\documentclass[a4paper, accepted=2025-02-02]{quantumarticle}
\pdfoutput=1
\usepackage[ascii]{inputenc}
\usepackage{amsmath,amssymb,amsfonts,amsthm}
\usepackage{mathtools}
\usepackage{braket}
\usepackage{notoccite}
\usepackage{dsfont}
\usepackage{circuitikz}
\usepackage{tikz}
\usepackage{soul}
\usepackage{graphicx}
\usetikzlibrary{quantikz}
\usepackage{tikz}
\usetikzlibrary{decorations.pathreplacing,decorations.pathmorphing}
\usepackage[caption=false]{subfig}
\usepackage{enumitem}
\usepackage[ruled]{algorithm2e}
\usepackage{hyperref}

\DeclarePairedDelimiter\ceil{\lceil}{\rceil}
\DeclarePairedDelimiter\floor{\lfloor}{\rfloor}

\renewcommand{\refname}{References}

\usepackage[numbers]{natbib}
\DeclareMathOperator*{\argmin}{arg\,min}

\DeclareMathOperator{\e}{e}
\DeclareMathOperator{\erf}{erf}

\DeclarePairedDelimiter\abs{\lvert}{\rvert}
\DeclarePairedDelimiter\norm{\lVert}{\rVert}

\newtheorem{theorem}{Theorem}

\newtheorem{definition}[theorem]{Definition}
\newtheorem*{proposition*}{Proposition}
\newtheorem{obj}{Objective}

\begin{document}

\title{Enhancing Scalability of Quantum Eigenvalue Transformation of Unitary Matrices for Ground State Preparation through Adaptive Finer Filtering}

\author{Erenay Karacan}
\email{ekaracan@ethz.ch}
\affiliation{Technical University of Munich, School of Computation, Information and Technology, Boltzmannstra{\ss}e 3, 85748 Garching, Germany}
\author{Yanbin Chen}
\email{yanbin.chen@tum.de}
\affiliation{Technical University of Munich, School of Computation, Information and Technology, Boltzmannstra{\ss}e 3, 85748 Garching, Germany}
\author{Christian B.~Mendl}
\email{christian.mendl@tum.de}
\affiliation{Technical University of Munich, School of Computation, Information and Technology, Boltzmannstra{\ss}e 3, 85748 Garching, Germany}
\affiliation{Technical University of Munich, Institute for Advanced Study, Lichtenbergstra{\ss}e 2a, 85748 Garching, Germany}

\begin{abstract}
Hamiltonian simulation is a domain where quantum computers have the potential to outperform their classical counterparts. One of the main challenges of such quantum algorithms is increasing the system size, which is necessary to achieve meaningful quantum advantage. In this work, we present an approach to improve the scalability of eigenspace filtering for the ground state preparation of a given Hamiltonian. Our method aims to tackle limitations introduced by a small spectral gap and high degeneracy of low energy states. It is based on an adaptive sequence of eigenspace filtering through Quantum Eigenvalue Transformation of Unitary Matrices (QETU) combined with spectrum profiling. By combining our proposed algorithm with state-of-the-art phase estimation methods, we achieved good approximations for the ground state energy with local, two-qubit gate depolarizing probability up to $10^{-4}$. To demonstrate the key results in this work, we ran simulations with the transverse-field Ising Model on classical computers using \texttt{Qiskit}. We compare the performance of our approach with the static implementation of QETU and show that we can consistently achieve three to four orders of magnitude improvement in the absolute error rate.

\end{abstract}

\maketitle
\section{Introduction}
A natural application field of quantum computers is calculating a given Hamiltonian's ground state and the corresponding energy. This task is pivotal in practical application fields, such as material science  \cite{Lesar2013-yx}, quantum chemistry \cite{dirac_quantum_chemistry, Liu_2022} and condensed matter physics \cite{cmp_applications1, cmp_applications2}. Moreover, a general computational problem can be mapped into calculating the ground state energy of a designated Hamiltonian \cite{Kitaev2002}, indicating the universal utility of ground state preparation. In light of this importance, various quantum algorithms have been proposed to efficiently prepare ground states of paradigmatic Hamiltonians \cite{McArdle_2020, Bauer_2020}. One particularly promising family of approaches, that scale near-optimally with respect to the spectral gap of the system, is based on eigenspace filtering \cite{kitaev1995quantummeasurementsabelianstabilizer, Poulin_2009, Lu_Cosine, Irmejs_Filter, Lin_2020_filtering}. In the adiabatic ground state preparation, total evolution time scales as $\mathcal{O}\left(\Delta_{\text{min}}^{-2} \right)$ for $\Delta_{\text{min}}$ being the minimal spectral gap along the evolution \cite{Born27, Farhi_AQC, Jansen_AQC, Albash_AQC}. In contrast to that, it was recently shown that we can achieve near-optimal scaling of $\mathcal{O}\left(\Delta^{-1} \log(\Delta^{-1}) \right)$ for $\Delta$ being the spectral gap of the Hamiltonian, through the so-called ``Quantum Eigenvalue Transformation of Unitary Matrices'' (QETU) ~\cite{qetu} that employs a polynomial filter approximating a step function via quantum signal processing (QSP) \cite{Martyn_2021}. There, the Hamiltonian block-encoding was substituted by the unitary time evolution operator, which is likely more straightforward to realize (via Trotterization) than the conventional block-encoding, particularly in the early fault-tolerant regime. This algorithm works well for systems where the following three assumptions are fulfilled: 1) The spectral gap $\Delta$ between the ground and first excited state is large enough such that the polynomial degree needed in QETU is in the reasonable range (below $\sim$ 50) 2) A cut-off value $\mu$ that bisects the ground state energy and first excited state energy can be guessed or estimated accurately 3) The overlap of the initial state with the ground state is large enough to achieve sensible success probabilities. Although starting with a high enough initial overlap remains a prerequisite for most existing algorithms, here we focus on the first two assumptions, which are difficult to fulfill as the system size increases, because the spectral gap $\Delta$ is expected to decrease for larger systems. In this situation, the QETU algorithm not only requires a correspondingly higher polynomial degree of the QSP sequence to have a sharp enough transition, but also necessitates a much more precise cut-off value $\mu$. 

\begin{figure*}
\centering
\begin{quantikz}[thin lines] 
            \lstick{$\ket{0}$}&  \gate{e^{i\phi_0X}} & \ctrl{1} & \gate{e^{i\phi_1X}}& \ctrl{1} &\qw\hspace{1mm} \dots \hspace{1mm}  &\ctrl{1} & \gate{e^{i\phi_1X}}  & \ctrl{1} & \gate{e^{i\phi_0X}} & \meter{} \\  
            \lstick{$\ket{\psi}$} & \qw & \gate{U} & \qw & \gate{U^{\dag}}  &\qw\hspace{1mm} \dots \hspace{1mm}  &  \gate{U} & \qw & \gate{U^{\dag}} & \qw &  \frac{F(\cos(H/2)) \ket{\psi}}{ \norm{ \bra{\psi}F(\cos(H/2)) \ket{\psi}}}
\end{quantikz}
\caption[Quantum Eigenvalue Transformation of Unitary Matrices (QETU) Circuit]{ 
Quantum Eigenvalue Transformation of Unitary Matrices (QETU) Circuit in compact notation where $U$ is the multi-qubit gate applying the time evolution operator $U=e^{-iH}$, acting on all the system qubits and the X-Rotation gates are applied to the ancilla qubit. Symmetric phases $(\phi_0, \phi_1, \dots \phi_1, \phi_0) \in \mathds{R}^{\eta+1}$ are optimized for a given target polynomial $F(a)$. 
} \label{fig:qetu_circuit}
\end{figure*}
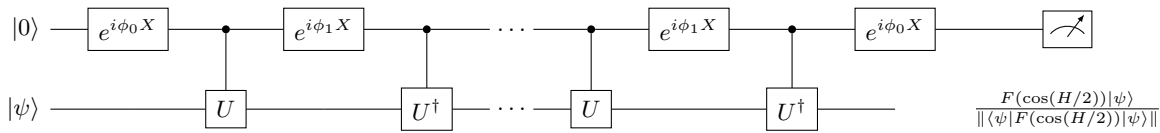

In this work, we propose an adaptive set of finer filtering stages to alleviate the challenges introduced by the small spectral gap with a trade-off over increased total simulation time, where we keep the polynomial degree fixed. In order to benchmark the performance of our proposed approach, we ran a series of numerical simulations of our proposed algorithm, as well as the static version of QETU, where we employed the Riemannian quantum circuit optimization (RQC-Opt) \cite{rqcopt} algorithm for compressing the circuit implementation of the time evolution operator and ran the algorithms under different noise levels. Our results demonstrate that our approach can yield a final state whose overlap with the ground state is larger compared to the static repetition of QETU. When combined with efficient phase estimation methods, such as the Robust Phase Estimation algorithm proposed by Ni et al.~\cite{rpe} or the Quantum Complex Exponential Least Squares (QCELS) algorithm proposed by Ding et al.~\cite{qcels}, we observe significant improvement in the approximation error compared to the static variant.

The main idea of our work is based on intermediate ``stretching'' of the Hamiltonian after each stage of applying QETU to achieve successively finer filtering. This idea necessitates estimating how much of the spectrum could be successfully filtered in the previous stages so that unwanted amplification of higher energy states can be avoided after the stretching. For this purpose, we employ the algorithm proposed by Lin and Tong \cite{LT}. This algorithm aims to approximate the Cumulative Distribution Function (CDF) of an input state (with respect to the given Hamiltonian) by employing the Fourier series approximation of the periodic error function in combination with its Fourier moments, which are to be acquired through quantum simulations. We use this approach in our ``profiling'' stages to determine an updated spectrum length after each successful filtering.

\section{Background}
\subsection{Fundamentals of QETU}
Quantum Signal Processing (QSP) framework offers a promising unification of quantum algorithms, \cite{Martyn_2021} where one can manipulate probability amplitudes of a subset of qubits as a function of certain circuit parameters (encoded in the so-called ``signal operator''), through tailoring another parameter set of the circuit, referred to as: ``QSP phases''. This manipulation happens through a targeted polynomial, for which one optimizes the QSP phases. As shown in \cite{pyqsp}, this optimization can be achieved up to machine precision efficiently and there is a one-to-one correspondence with a targeted polynomial and corresponding QSP phases. Main idea behind the unifying character of this proposal is based on approximating any arbitrary signal processing operation through a polynomial and carefully choosing the signal operator to represent the system or computational task we have.

The authors of \cite{qetu} extend this framework to show that we can use QSP to perform eigenspace filtering, where they use the controlled time evolution operator of the target Hamiltonian as the signal operator and optimize the QSP phases for a real, even polynomial $F(a)$. The corresponding circuit sequence can be written as:
\begin{equation}
    \mathcal{U}_{\text{QETU}} = e^{i \phi_0 X} cU^{\dag} e^{i \phi_1 X} cU \dots e^{i \phi_1 X} cU e^{i \phi_0 X}
\end{equation}
where $cU$ is the controlled time evolution operator ($U = e^{-iH}$ with the time step omitted into $H$) and $\Vec{\phi} = (\phi_0, \phi_1, \dots, \phi_1, \phi_0)$ are the QSP phases.

As stated in \cite[Theorem 1]{qetu} and visualized in Fig.~\ref{fig:qetu_circuit}, the QETU circuit with these symmetric phases $\Vec{\phi} \in \mathbb{R}^{\eta+1}$, optimized for a target polynomial $F(a)$ and applied to a given input state $\ket{\psi}$ delivers the following final state, upon post-selecting the ancilla qubit in $\ket{0}$:
\begin{equation}
\label{eq:qetu_algebraic}
\begin{split}
&\bra{0}_{\text{anc}} \mathcal{U}_{\text{QETU}} \ket{0}_{\text{anc}} \ket{\psi}
= \frac{F(\cos(\frac{H}{2})) \ket{\psi}}{ \norm{ \bra{\psi}F(\cos(\frac{H}{2})) \ket{\psi}}}\\
&= \frac{1}{ \norm{ \bra{\psi} F(\cos(\frac{H}{2})) \ket{\psi}}} \sum_{j} c_j F(\cos(\lambda_j/2)) \ket{\psi_j},
\end{split}
\end{equation}
where $\{\lambda_j, \ket{\psi_j} \}_j$ is the eigensystem of the Hamiltonian and $c_j = \braket{\psi \vert \psi_j}$ is the overlap of the initial state with the eigenstate $\ket{\psi_j}$. Here, $F(a)$ has to be real-valued, with parity $\eta \mod 2$, maximum degree $\eta$ and has to satisfy $\abs{F(a)} \leq 1$ for all $a \in [-1, 1]$. 

\subsection{Eigenspace Filtering with QETU}
\label{sec: eigenspace_filtering}
\begin{figure}[htb]
  \centering
  \includegraphics[scale=0.55]{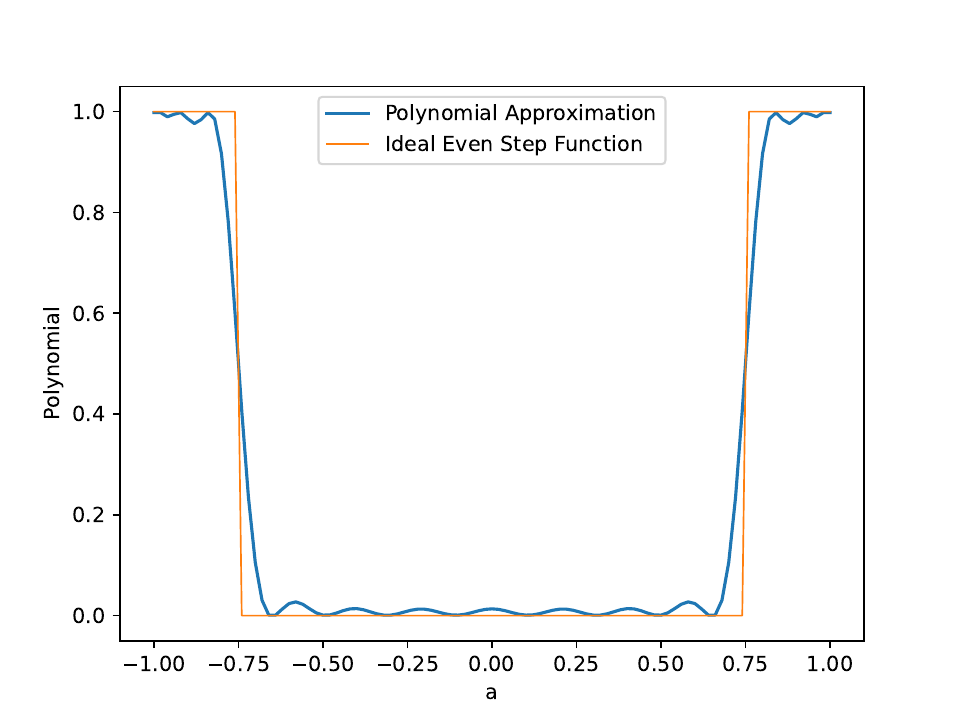}
  \caption{Example polynomial $F(a)$ of degree 30, as an approximation of the even step function with cut-off value $\mu = 0.75$. We used Chebyshev polynomials of the second kind as basis and employed the convex optimization library ``cvxpy'' to optimize coefficients of the Chebyshev polynomials.} \label{fig:step}
\end{figure}

With sensible estimates for the lower and upper bounds of the spectrum ($\lambda_{\text{LB}}, \lambda_{\text{UB}}$), we can ensure that the linearly transformed eigenspace is in the interval $(0, \pi)$. The linear transformation can be applied as:
\begin{equation}
\label{eq:linear_trafo}
\Tilde{H} \coloneqq \frac{\pi (H - \lambda_{\text{LB}} I)}{\lambda_{\text{UB}} - \lambda_{\text{LB}}}.
\end{equation}
Then, the cosine transformation is strictly decreasing and maps the spectrum to $(0, 1)$.
The overall transformation implies the following relation between the transformed $a \coloneqq \cos(\frac{\Tilde{\lambda}}{2})$ space and $\lambda$ space:

\begin{equation}
a_j \coloneqq \cos\left(\frac{\pi}{2} \frac{(\lambda_j - \lambda_{\text{LB}})}{\lambda_{\text{UB}} - \lambda_{\text{LB}}}\right)
\end{equation}
where $\{\lambda_j\}_j$ is the spectrum of the Hamiltonian $H$.
We now introduce the main task as follows:
\begin{obj}[Ground State Preparation]
    \label{obj:gsp}
    Suppose we are given a Hamiltonian $\Tilde{H}$, whose spectrum is contained in (0, $\pi$), whose first excited state energy and ground state energy are separated by a spectral gap $\Delta > 0$ and which can be accessed through its time evolution operator; the goal is to find an approximation $\ket{\psi_f}$ of the ground state $\ket{\psi_0}$, such that $\abs{\braket{\psi_0|\psi_f}} \geq 1-\epsilon$.
\end{obj}
Dong et al.~\cite{qetu} show that if we employ a QETU circuit where $\Tilde{H}$ is used in the time evolution operator and optimize the QSP phases for a polynomial $F(a)$ with a step-like transition behavior from 0 to 1 at the cut-off value $\mu \in (a_1, a_0)$, we can amplify the overlap of any initial arbitrary state with the ground state and dampen the overlap with all excited states. An example of such a polynomial $F(a)$ is given in Fig. \ref{fig:step}.

In principle, any polynomial that is monotonously increasing in the (0, 1) range can be used to amplify the ground state relative to excitations, as long as we are sure of the bijective mapping of the spectrum to the (0, $\pi$) interval. This relative amplification of the ground state (which we will formally define in Def. \ref{def:relative_ampf}) depends significantly on the polynomial degree and spectral gap. As shown in \cite[Theorem 6]{qetu}, this necessitates that we scale the polynomial degree $\eta$, hence also the query depth of the time evolution operator as:
\begin{equation}
    \label{eq:qetu_query_depth}
    \Tilde{\mathcal{O}}(\Delta^{-1} \log(\epsilon^{-1})) \text{ queries to the (controlled)-U. }
\end{equation}

This, however, implies that we have to arbitrarily increase the polynomial degree for an arbitrarily small spectral gap $\Delta$ and/or error $\epsilon$. In turn, this increases the circuit depth significantly. An alternative, more efficient way is to apply QETU with (relatively) lower degree multiple times. This cuts down on the circuit depth, as the longest path is dominated by the ancilla qubit and this qubit is reset between each repetition of QETU. Hence, throughout the rest of the paper we retain a fixed polynomial degree. Moreover, we define a new set of parameters and metrics to assess the performance of this algorithm.
Equivalent to increasing the polynomial degree arbitrarily in this way, one can repeat the QETU circuit of fixed degree and achieve the same scaling, as shown in Appendix \ref{sec:A}. In the following, we will retain a fixed polynomial degree and benchmark our proposed approach with this static repetition variant. Moreover, we will introduce some performance metrics to assess the performance of both variants (static repetition and our proposed approach).

Due to the filtering nature of QETU, we will focus on the magnitude of amplification each repetition of the circuit induces to the ground state overlap, relative to excited state overlaps. This is a particularly useful metric, as it can be directly related to the polynomial $F(a)$ (see Eq.~\eqref{eq:qetu_algebraic} and Appendix \ref{sec:A} for more details), and it also gives us a direct correspondence with the final ground state overlap. Moreover, we can limit our consideration to the relative amplification between the ground state and the first (non-negligibly occupied) excitation overlap, if we assume $F(a)$ is monotonously increasing in $a \in (0, 1)$ and the spectrum was successfully mapped to $\Tilde{\lambda} \in (0, \pi)$ bijectively (see Eq.~\eqref{eq:linear_trafo}), as this indicates $F(a_j) < F(a_1), \hspace{.4cm} \forall j > 1$.

\begin{definition}[Relative Amplification]
\label{def:relative_ampf}
The relative amplification between an initial state $\ket{\psi_{\text{init}}}$ and a final state $\ket{\psi_f}$, in relation to the ground and first excited states, is defined as:
\begin{equation}
    A \coloneqq \sqrt{\frac{\sum_{j}|\braket{\psi_f|\psi_{0,j}}|^2/\sum_j |\braket{\psi_{\text{init}}|\psi_{0, j}}|^2}{\sum_{j}|\braket{\psi_f|\psi_{1,j}}|^2/\sum_j |\braket{\psi_{\text{init}}|\psi_{1, j}}|^2}}
\end{equation}
where $\text{span}\{\ket{\psi_{0, j}}\}_j, \text{span}\{ \ket{\psi_{1, j}}\}_j$ are the ground and first excited eigenspaces, respectively.
\end{definition}
In this definition we allowed degeneracy in the ground and first excited states, where the vectors $\ket{\psi_{0, j}}$ and $\ket{\psi_{1, j}}$ are the degenerate eigenvectors. The relative amplification can be used as a strong indicator of state fidelity. If $A \approx 1$, the relation between the ground and first excited state overlaps remains approximately the same after the amplification. For $A \rightarrow \infty$ we can infer that $\abs{\braket{\psi_f | \psi_1}} \rightarrow 0$. This also indicates that for all other excited states $\abs{\braket{\psi_f | \psi_j}} \rightarrow 0, \forall j>1$. This results from the monotonously increasing nature of $F(a)$, for the $a$ values mapped into the step-like transition interval and we assume that outside of the transition interval $F(a)$ oscillates with a negligible amplitude. Hence,
\begin{equation}
    \label{eq:A_to_inf}
    A \rightarrow \infty \Longrightarrow \abs{\braket{\psi_f | \psi_0}}  \rightarrow 1.
\end{equation}
With this performance metric, we will re-formulate the main objective with an error parameter $\Tilde{\epsilon}$, called ``inverse amplification strength''.
\begin{obj}[Ground State Overlap Amplification]
\label{def:gsa}
With the same assumptions from Obj.~\ref{obj:gsp} and Def.~\ref{def:relative_ampf}, the goal is to prepare a quantum state $\ket{\psi_f}$ with a relative amplification $\Tilde{A}$, that satisfies: 
\begin{equation}
    \label{eq:relative_ampf_cond}
    \Tilde{A} = \frac{a_0}{\Delta_a \Tilde{\epsilon}}
\end{equation}
where  $\Tilde{\epsilon} \ll 1$ is the ``inverse amplification strength'' and $\Delta_a \coloneqq a_0 - a_1$ is the spectral gap in the cosine space.
\end{obj}
Inverse Amplification Strength $\Tilde{\epsilon}$ can be taken as an indicator for the final state infidelity $\epsilon$ (defined in Obj.~\eqref{obj:gsp}). This comes from the fact that the relative amplification is directly related to the end ground state overlap. If $\Tilde{\epsilon} \rightarrow 0$, then $A \rightarrow \infty$ and also $\epsilon \rightarrow 0$. Due to this behaviour, we will use this parameter to represent the error of our main objective throughout the rest of this article.

\begin{definition}[Simulation Time]
    \label{def:t}
    The simulation time of a Hamiltonian simulation circuit is defined as:
    \begin{equation}
        T = \sum_j t_j
    \end{equation}
    where $\{t_j\}_j$ are the time steps of each time evolution operator $U_j = \e^{-i H t_j}$ employed in the quantum circuit. 
\end{definition}

\begin{definition}[Maximal Simulation Time]
    \label{def:tmax}
    The maximal simulation time of a Hamiltonian simulation algorithm is defined as:
    \begin{equation}
        T_{\text{max}} = \max \{T_1, T_2, \dots\}
    \end{equation}
    where $\{T_j\}_j$ are the simulation times of the respective quantum circuits employed in different stages of the algorithm. 
\end{definition}

\begin{definition}[Total Simulation Time]
    \label{def:ttot}
    Total simulation time of a Hamiltonian simulation algorithm is defined as:
    \begin{equation}
        T_{\text{total}} = \sum_j T_j
    \end{equation}
    where $\{T_j\}_j$ as given in Def. \ref{def:tmax}. 
\end{definition}

We defined this ``maximal'' simulation time to represent the deepest quantum circuit that has to be accommodated by the hardware. We distinguish maximal from ``total'' simulation time, where total simulation time includes simulation times from the stages of the algorithm with shallower circuit depths than the circuit sequence with the most gate operations.

Eventually, we want to assess the performance of our algorithms by comparing the scaling of the (maximal) circuit depth required to implement them on the quantum hardware. As there is a linear correspondence between simulation time and the number of gate operations required through optimal Hamiltonian simulation \cite{suzuki1976generalized, suzuki1985general}, we want to define the circuit depth parameter in linear relation to $T_{\text{max}}$. Moreover, because we focus on the number of repetitions of a QETU sequence of fixed polynomial degree, we want to define the circuit depth relevant for our purposes independent of the polynomial degree $\eta$. Hence, we formalize the circuit depth definition through:
\begin{definition}[Circuit Depth]
    \label{def:circuit_depth}
    Circuit depth required for a ground state overlap amplification process (using QETU with polynomial $F(a)$ and spectrum bounds $\lambda_{\text{LB}}, \lambda_{\text{UB}}$) is given as the constant factor $\gamma$, which scales the maximal simulation time $T_{\max}$ as:
    \begin{equation}
        T_{\max} = \gamma \eta \frac{\pi}{\lambda_{\text{UB}} - \lambda_{\text{LB}}}
    \end{equation}
    where $\eta$ is the degree of the polynomial $F(a)$.
\end{definition}

Here $\frac{\pi}{\lambda_{\text{UB}} - \lambda_{\text{LB}}}$ term represents the (basic) time step that has to be used in each (controlled)-U operator in order to perform the linear transformation of the Hamiltonian, as given in Eq.~\eqref{eq:linear_trafo}.  For the static repetition of QETU, where spectrum is mapped to $(0, \pi)$ at every filtering operation, $\gamma$ can be seen as the number of circuit repetitions. As we will see in Section \ref{sec:AFF}, if we decide to perform a stretching to the Hamiltonian (hence also a different spectrum mapping), we will include this time step scaling factor in $\gamma$. In this case $\gamma$ cannot be treated as the integer number of circuit repetitions anymore but will also be dependent on the stretching factor.

\begin{proposition*}
\label{prop:static_depth}
A final state $\ket{\psi_f}$ with relative amplification $\Tilde{A}$ (satisfying Eq.~\eqref{eq:relative_ampf_cond}) can be achieved by applying QETU on an initial state $\ket{\psi_{\text{init}}}$ repeatedly $\ceil*{\gamma_{\text{ST}}}$ times, where circuit depth $\gamma_{\text{ST}}$ (as defined in Def.~\ref{def:circuit_depth}) scales as:
\begin{equation}
    \label{eq:static_depth}
    \gamma_{\text{ST}} = \mathcal{O}(\Delta^{-1} \log(\Delta^{-1} \Tilde{\epsilon}^{-1}))
\end{equation}
where $\Delta = \lambda_1 - \lambda_0$ is the spectral gap and $\Tilde{\epsilon}$ is the inverse amplification strength. 
\end{proposition*}
For the proof of the proposition, please refer to Appendix~\ref{sec:A}.

It is also worth stating that the initial overlap problem persists with this method, manifesting itself as causing a low success probability for the amplification, if the overlap of the initial state with the ground state is low. In our simulations, we randomize the initial state and limit the number of measurements performed on the ancilla qubit proportional to the success probability of each ancilla qubit being post-selected as $\ket{0}$. A much more effective method could be using another quantum oracle to achieve an initial state whose overlap with the ground state is increased compared to randomized initialization.

\subsection{Spectrum Profiling}
So far we have reviewed the QETU sequence and its usage for spectrum filtering, with the goal of preparing a high fidelity ground state through a set of amplifications through a polynomial filter. Now we will summarize a potential method to perform a ``spectrum profiling''. This idea is pivotal in our proposed approach as it is thanks to this stage we will be able to better estimate a cut-off value $\mu$ for the polynomial and successively better the choice of spectrum bounds, which effectively means we will use it to stretch the Hamiltonian in the time evolution operator as we filter out excitations. As proposed by Lin \& Tong \cite{LT}, given the Hamiltonian $H$ with the spectrum $\{\lambda_j, \ket{\psi_j}\}_j$, the Cumulative Distribution Function (CDF) of the input state $\ket{\psi_{\text{init}}}$ is defined in the following way:
\begin{equation}
    C(x) \coloneqq \sum_{j: \lambda_j \leq x} \left|\braket{\psi_{\text{init}} | \psi_j}\right| .
\end{equation}
If we have access to lower and upper bounds for the spectrum $(\lambda_{\text{UB}}, \lambda_{\text{LB}})$, hence also to an upper bound for the spectrum length $\Lambda \coloneqq \lambda_{\text{UB}} - \lambda_{\text{LB}}$, we can approximate (a periodic repetition of) CDF as the convolution of the $\Lambda$-periodic continuation of the error function $\erf_{\Lambda}(x)$ and the spectral density $p(x) \coloneqq \sum_j \left|\braket{\psi_{\text{init}} | \psi_j}\right| \delta(x - \lambda_j)$:
\begin{equation}
    C(x) \approx (\erf_{\Lambda} \ast \hspace{0.05cm} p)(x), \hspace{0.5cm} \forall x \in (\lambda_{\text{LB}}, \lambda_{\text{UB}}).
\end{equation}
Furthermore, we can approximate $\erf_{\Lambda}$ by a Fourier series expansion with (2D+1)-terms:
\begin{equation}
    \erf_{\Lambda}(x) \approx \sum_{|k| \leq D} F_k \e^{ikx},
\end{equation}
where $F_k$ are the Fourier coefficients.
From this, it follows that:
\begin{equation} \label{eq:fourier_moments}
    C(x) \approx \sum_{|k|\leq D} F_k \e^{ikx} \braket{\psi_{\text{init}}|\e^{-i k H}|\psi_{\text{init}}} 
\end{equation}
where the Fourier moments $\braket{\psi|\e^{-i k H}|\psi}$ can be efficiently acquired through quantum simulation by employing a simple Hadamard test circuit, as shown in Fig.~\ref{fig:hadamard_circuit}.
For the case $\Lambda = 2$, the Fourier coefficients read \cite{kiss2024early}:
\begin{subequations}
\label{eq:fourier_coeffs}
\begin{equation}
    \label{eq:fourier_coeffs1}
    F_0 = \frac{1}{2}
\end{equation}
\begin{equation}
    \label{eq:fourier_coeffs2}
    F_{2j+1} = -i \sqrt{\frac{\beta}{2\pi}} \e^{-\beta} \frac{I_j(\beta) + I_{j+1}(\beta)}{2j+1}
\end{equation}
\begin{equation}
    \label{eq:fourier_coeffs3}
    F_{D} = -i \sqrt{\frac{\beta}{2\pi}} \e^{-\beta} \frac{I_{\floor*{\frac{D-1}{2}}}(\beta)}{D}
\end{equation}
\end{subequations}
where $j \in \{ \floor*{\frac{-D-1}{2}}, \dots , 0, \dots,  \floor*{\frac{D-3}{2}} \}$, $I_n$ is the n-th modified Bessel function of the first kind and $\beta$ to be chosen depending on the target approximation precision, with a trade-off of larger contribution from the higher order Fourier terms. It is important to note that $F_{2k} = 0$, hence we need to compute only $\floor*{\frac{D-1}{2}}$ Fourier moments through quantum simulation, which alleviates the total simulation time on the quantum hardware.

\begin{figure}[!h]
\centering
\begin{quantikz}[thin lines] 
    \lstick{$\ket{0}$}&  \gate{H} & \gate{I / S^{\dag}} & \ctrl{1} & \gate{H} & \meter{} \\  
    \lstick{$\ket{\psi}$} & \qw & \qw & \gate{\e^{-i k H}} & \qw & \qw
\end{quantikz}
\caption{Hadamard test circuit used to compute the Fourier moments in Eq.~\eqref{eq:fourier_moments}. The identity $I$ is inserted to compute $\text{Re}\braket{\psi|\e^{-i k H}|\psi}$ and the conjugated phase gate $S^{\dag}$ is used to compute $\text{Im}\braket{\psi|\e^{-i k H}|\psi}$.
} \label{fig:hadamard_circuit}
\end{figure}

\section{Adaptive Finer Filtering}
\label{sec:AFF}
The eigenspace filtering method we presented in the previous section faces multiple practical limitations when we study large systems and complicated Hamiltonians. One of which, is the difficulty of finding a cut-off value $\mu$ for an arbitrary input state and verifying that the spectrum bounds indeed result in the correct mapping of $\Tilde{\lambda} \in (0, \pi)$. Moreover, for large spectral range $\lambda_{\text{UB}} - \lambda_{\text{LB}}$, the transformation in Eq.~\eqref{eq:linear_trafo} results in a tightly spaced spectrum in $\Tilde{\lambda}$ space, which significantly limits the interval $\mu$ has to lie in and greatly reduces the relative amplification one can achieve for a given polynomial degree.
Hence, instead of relying on heuristics for the choice of $\mu$ and to make sure that our spectrum bounds are more efficient, we propose an algorithm called ``Adaptive Finer Filtering'' in this section. The main objective of our approach can be stated as follows:

\begin{obj}[Adaptive Finer Filtering (AFF)]
With the assumptions from Def. \ref{def:gsa}, we aim to achieve relative amplification $\Tilde{A}$ satisfying the condition in Eq.~\eqref{eq:relative_ampf_cond}, while the circuit depth $\gamma_{\text{AFF}}$ scales as:
\begin{equation}
    \label{eq:aff_depth}
    \gamma_{\text{AFF}} = \mathcal{O}(\Delta^{-1})
\end{equation}
in relation to the spectral gap $\Delta$, independent of amplification strength $\Tilde{\epsilon}^{-1}$.
\end{obj}

Our algorithm offers to be more efficient and more rigorous compared to the static repetition of QETU with unchanged spectrum bounds. It filters out excitations with decreasing spectrum length at each repetition - hence, operates through successive stretching of the Hamiltonian. For each repetition of QETU, we take only the non-negligibly occupied excited states into account in our $\lambda \rightarrow \Tilde{\lambda}$ mapping and thereby make sure that the highest excitation gets mapped to a range close to 0 in the $a = \cos(\Tilde{\lambda}/2)$ space. This intentional mapping not only alleviates the minimal polynomial degree $\eta$ required to have a sharp enough transition (hence also improving the circuit depth scaling) but also enables us to have a recipe for the choice of the cut-off $\mu$ at each stage.

To perform this improved $\lambda \rightarrow \Tilde{\lambda}$ mapping, we estimate new upper and lower bounds after each filtering stage through the aforementioned algorithm by Lin and Tong \cite{LT}. In this intermediate ''profiling stage'', updated spectrum bounds are taken as boundaries of the interval where the growth rate of the CDF is the highest. This is determined by the region where the first derivative of the CDF is greater than $\xi_1$ and the absolute value of the second derivative is smaller than $\xi_2$. Here $(\xi_1, \xi_2)$ are user-set parameters that can be determined depending on the acquired CDF.

A systematic summary of the Adaptive Finer Filtering Algorithm and its helper method for the profiling stage can be found in Algorithms \ref{alg:alg1} and \ref{alg:alg2}. To see how we achieve the target scaling in Eq.~\eqref{eq:aff_depth}, refer to Appendix~\ref{sec:proof2}. We also note that the scaling given in Eq.~\eqref{eq:aff_depth} achieves the optimal $\Delta$-scaling (as shown in \cite{Lin_2020}) but unlike the algorithm proposed in \cite{Lin_2020} assumes no lower bound for the spectral gap and employs unitary time evolution through Trotterization instead of block-encoding.

Algorithm \ref{alg:alg1} takes an initial state $\ket{\psi_{\text{init}}}$ as the input and delivers an approximation of the ground state $\ket{\psi^M}$ after a total of $M$ filtering operations. In between each stage, we perform a spectrum profiling (given in Algorithm \ref{alg:alg2}) to estimate the highest occupied excitation such that it can be used as the spectrum upper bound in the following next filtering stage, effectively stretching the Hamiltonian successively. Using the information extracted from spectrum profiling we also choose the cut-off value $\mu$ systematically at each stage. 

This algorithm has certain user-set parameters, such as the number of filtering stages $M$, polynomial degree for QETU $\eta$, initial spectrum bounds  $\lambda_{\text{LB}}^0, \hspace{0.1cm} \lambda_{\text{UB}}^0$ (which can be arbitrarily broad), precision bounds for the spectrum profiling $(\xi_1, \xi_2)$, number of Fourier terms $D$ to approximate the CDF (as given in Eq.~\eqref{eq:fourier_coeffs}) and division coefficients $m_i$, which are used to determine the cut-off values $\mu^i$ of each (i-th) filtering stage. These division coefficients are to be chosen, such that $\Tilde{\mu}^i \coloneqq 2\arccos(\mu^i)$ cuts the spectrum into two as:
\begin{equation}
    \Tilde{\mu}^i = c^i_1 \Big(\lambda_{\text{LB}}^i +  \frac{\Lambda^i}{m_i}\Big) + c^i_2, 
\end{equation}
where $\lambda^i_{\text{LB}}$ is the (estimated) lower bound of the (filtered) spectrum, $\Lambda^i$ is the (estimated) total length of the (filtered) spectrum and the $(c^i_1, c^i_2)$ coefficients are chosen according to the linear transformation shown in Eq.~\eqref{eq:linear_trafo}.

In order to reduce the possibility of $\mu$ being greater than $a_0$, one can choose $m_i$ values adaptively. This case ($\mu > a_0$) is initially unlikely, so one can choose a larger $m_i$ to cut off the spectrum into a smaller piece in early stages. In later stages, $m_i$ can be chosen as 2 to avoid overfiltering. 

Another input that Algorithm \ref{alg:alg1} takes is $\mu^0$, the first cut-off value of the filtering sequence and can be set to a value between $(0.9, 0.95)$, in practice. If the error margins for the estimates $\lambda_{\text{LB}}^0, \hspace{0.1cm} \lambda_{\text{UB}}^0$ (initial lower and upper bounds) are large, one can perform spectrum profiling with arbitrarily high/low bounds on the initial state before the first filtering stage to determine the bounds and $\mu^{0}$ more accurately.

\begin{algorithm}
\caption{Adaptive Finer Filtering for Ground State Preparation}\label{alg:alg1}
\KwData{$\ket{\psi_{\text{init}}}, \hspace{0.1cm} \lambda_{\text{UB}}^0, \hspace{0.1cm} \lambda_{\text{LB}}^0, \hspace{0.1cm} \mu^0, \hspace{0.1cm} \eta, \hspace{0.1cm} \Vec{m}, \hspace{0.1cm} M, \hspace{0.1cm} \xi_1, \hspace{0.1cm} \xi_2, \hspace{0.1cm} D$}
{$\textbf{Ensure: } \forall \lambda_j \in [\lambda_{\text{LB}}^0, \lambda_{\text{UB}}^0]$; \hspace{0.1cm} $\eta$ is even}; \\ $\Vec{m}$ = ($m_0, \dots, m_{M-1}$). \\
$\ket{\psi^{0}} \gets \ket{\psi_{\text{init}}}$\\
\For{$i= 0 \dots M-1$} {
    $\ket{\psi^{(i+1)}} \gets \text{QETU}(\ket{\psi^{i}}, \mu^{i}, \eta, \lambda^{i}_{\text{LB}}, \lambda^{i}_{\text{UB}})$ \tcp{$\text{With  transformation as in Eq.~\eqref{eq:linear_trafo}}$}
    $\lambda^{(i+1)}_{\text{LB}}, \lambda^{(i+1)}_{\text{UB}} \gets \text{SP}(\ket{\psi^{(i+1)}}, \lambda^{i}_{\text{LB}}, \lambda^{i}_{\text{UB}}, \xi_1, \xi_2, D)$ \tcp{$\text{Algorithm \ref{alg:alg2}}$}
    $\Lambda^{(i+1)} \gets \lambda^{(i+1)}_{\text{UB}} - \lambda^{(i+1)}_{\text{LB}}$\\
    $c^{(i+1)}_1 \gets \frac{\pi}{2\Lambda^{(i+1)}}$\\
    $c^{(i+1)}_2 \gets -c^{(i+1)}_1 \lambda^{(i+1)}_{\text{LB}}$\\
    $\mu^{(i+1)} \gets \cos(c^{(i+1)}_1(\lambda^{(i+1)}_{\text{LB}} + \frac{\Lambda^{(i+1)}}{m_i}) + c^{(i+1)}_2)$
}
\Return $\ket{\psi^{M}}$
\end{algorithm}

\begin{algorithm}
\caption{Spectrum Profiling (SP)}\label{alg:alg2}
\KwData{$\ket{\psi}, \hspace{0.1cm} \lambda_{\text{UB}}, \hspace{0.1cm} \lambda_{\text{LB}}, \hspace{0.1cm} \xi_1, \hspace{0.1cm} \xi_2, \hspace{0.1cm} D$}
{$\textbf{Ensure: } \braket{\psi_j|\psi} \approx 0, \hspace{0.2cm} \forall \lambda_j \notin [\lambda_{\text{LB}}, \hspace{0.1cm} \lambda_{\text{UB}}]; \hspace{0.2cm} \xi_1, \xi_2 > 0$}\\
$\Lambda \gets \lambda_{\text{UB}} - \lambda_{\text{LB}}$\\
$\Tilde{H} \gets \frac{2}{\Lambda} (H - \lambda_{\text{LB}}I) - I \hspace{0.5cm}$ \tcp{$\textbf{Ensures } \Tilde{\lambda} \in (-1, 1)$}
$x \gets (-1, \dots , 1) \hspace{1.55cm}$ \tcp{$\text{energy grid}$}
$C(x) \gets \sum_{k=-D}^{D} F_k \e^{ikx} \braket{\psi|\e^{-ik\Tilde{H}}|\psi} \hspace{1cm}$ \tcp{Through Hadamard tests and Eqs.~\eqref{eq:fourier_coeffs}}
$C'(x) \gets \frac{d}{dx} C(x)$\\
$C''(x) \gets \frac{d}{dx} C'(x)$\\
$(x_{\text{LB}}, x_{\text{UB}}) \gets (x_1, x_2),$\\ $\hspace{0.6cm} \text{ s. t. } C'(x)>\xi_1 \hspace{0.2cm} \wedge \hspace{0.2cm} |C''(x)| < \xi_2, \hspace{0.2cm} \forall x \in [x_1, x_2]$ \\
\vspace{0.2cm}
\Return $\frac{\Lambda}{2} (x_{\text{LB}} + \frac{2\lambda_{\text{LB}}}{\Lambda} + 1)$, $\frac{\Lambda}{2} (x_{\text{UB}} + \frac{2\lambda_{\text{LB}}}{\Lambda} + 1)$
\end{algorithm}

\section{Simulation Results}
To investigate the performance of our approach, we ran a set of simulations using IBM \texttt{qiskit}. The source code of our implementation is available \href{https://github.com/erenaykrcn/AFF}{here}.

\begin{figure*}
\centering
\includegraphics[scale=0.5]{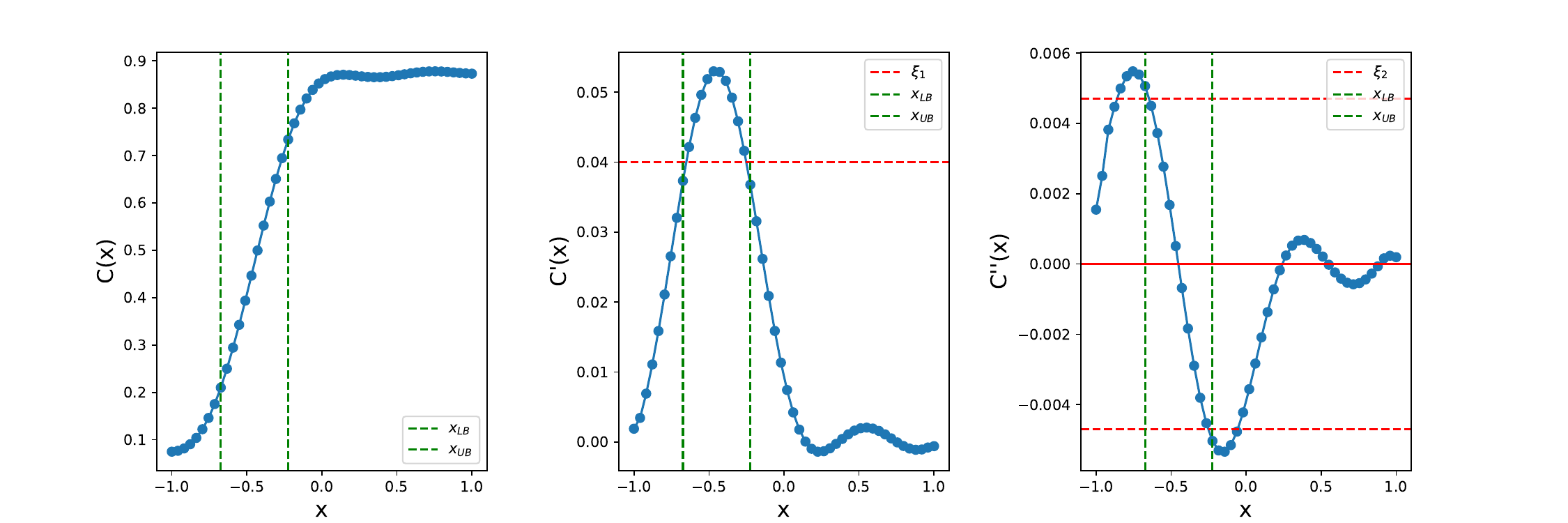}
\includegraphics[scale=0.5]{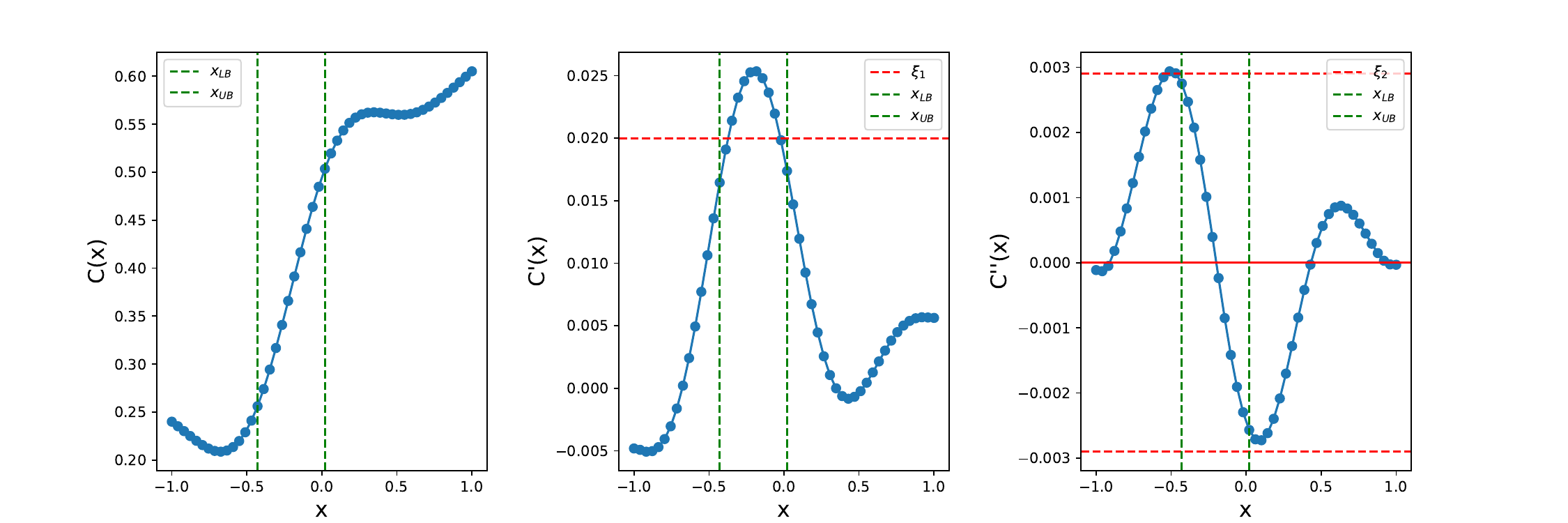}
\caption{Example plots of cumulative distribution functions (CDF) from the first (top) and second (bottom) filtering stages, plotted next to its first and second derivatives. Fourier moments were acquired through Hadamard sampling (Fig.~\ref{fig:hadamard_circuit}), using $10^3$ shots and simulated with a local depolarizing probability of $10^{-3}$ for the two-qubit gates. Red horizontal lines represent precision bounds for the spectrum profiling $\xi_1, \xi_2$, and green vertical lines represent new upper/lower spectrum bounds resulting from the spectrum profiling. The Hamiltonian is the TFIM with system size $L=6$ and parameters $J=1$, $g=1$. For the top graph: $(D, \beta, \lambda_{\text{LB}}, \lambda_{\text{UB}}) = (7, 5, -10, 10)$ and the input state $\ket{\psi}$ was taken as the result of the initial filtering by QETU with $\eta = 14$, $\mu = 0.95$, applied to a randomized state $\ket{\psi_{\text{init}}}$. For the bottom graph: $(D, \beta, \lambda_{\text{LB}}, \lambda_{\text{UB}}) = (7, 5, -8.78, -1.43)$ and the input state $\ket{\psi}$ was taken as the result of the secondary filtering by QETU with $\eta = 14$, $\mu = 0.92$, applied to the result of the initial filtering.}
\label{fig:cdf_plots}
\end{figure*}

We primarily investigated the (integrable) transverse-field Ising Model (TFIM) governed by the Hamiltonian
\begin{equation}
H_{\text{TFIM}} = -J\sum_{j=1}^{L-1} Z_j Z_{j+1} - g \sum_{j=1}^{L} X_j.
\end{equation}
Moreover, we implemented the QETU circuit with the control-free implementation of the time evolution circuit, as demonstrated in \cite[Section~VI]{qetu}.

For encoding the time evolution operator, we employed classical optimization through RQC-Opt \cite{rqcopt}. This toolbox uses the Riemannian trust region algorithm to optimize the two-qubit gates in a brick wall circuit layout to approximate the time evolution operator of a given Hamiltonian. A central idea of this approach is to optimize the gates for smaller systems and use these same gates to represent the time evolution operator of a larger system, assuming translation invariant Hamiltonians. Particularly for the TFIM model, it reaches good approximations with relatively short circuit depths.
In this work, we set the Hamiltonian coefficients as $J = 1$ and $g = 1$.

Heuristically, we fix the $(D, \beta, \xi_1, \xi_2)$ parameters to $(7, 5, 0.03, 0.02)$ for system sizes $L = 6, 8, 10$.
It is important to adjust these parameters for different system sizes and Hamiltonian parameters, depending on the required sharpness of the error function (determined by $D$ and $\beta$) and the values of $C'(x), C''(x)$ (for $\xi_1$ and $\xi_2$). Remarkably, we can get a spectrum profile for this choice of low $D$ by running the Hadamard sampling given in Fig.~\ref{fig:hadamard_circuit} only four times (for each time we execute Alg.~\ref{alg:alg2}). Namely, for even $k$, $F_k = 0$, and for negative $k$, we can conjugate the Fourier moment estimated for $-k$.

Example plots for $C(x)$ and its derivatives $C'(x), C''(x)$ are visualized in Fig.~\ref{fig:cdf_plots}. There one can spot the interval where the growth rate of the CDF is large. This interval is determined by the local maximum of the first derivative. We also demonstrate the amplification process in Table~\ref{tab:amplification} for the system sizes $L \in \{6, 8\}$ and $J=1$, $g=1$.

\begin{table}[]
\centering
\text{Ground State Preparation} \\
\vspace{0.3cm}
$L = 6; \hspace{0.5cm} \lambda_0 \approx -7.7274$ \\
\vspace{0.2cm}
\begin{tabular}{||c c c c c c||} 
     \hline
     $i$ & $\lambda_{\text{LB}}$ & $\lambda_{\text{UB}}$ & $\mu$ & $p$ & $\abs{\braket{\psi_0|\psi_f}}$ \\ [0.5ex] 
     \hline\hline
     0 & -10 & $10$  & $0.95$ & $0.053$ & $0.243$ \\ 
     \hline
     1 & -8.78 & $-1.43$  & $0.92$ & $0.362$ & $0.568$ \\ 
     \hline
     2 & -8.17 & $-6.98$  & $0.7$ & $0.502$ & $0.996$ \\ 
     \hline
     \hline
\end{tabular}\\
\vspace{0.5cm}
$L = 8; \hspace{0.5cm} \lambda_0 \approx -10.251 $ \\
\vspace{0.2cm}
\begin{tabular}{||c c c c c c||} 
     \hline
     $i$ & $\lambda_{\text{LB}}$ & $\lambda_{\text{UB}}$ & $\mu$ & $p$ & $\abs{\braket{\psi_0|\psi_f}}$ \\ [0.5ex] 
     \hline\hline
     0 & -15 & $15$  & $0.95$ & $0.007$ & $0.289$ \\ 
     \hline
     1 &  -12.55 & $-5.2$  & $0.7$ & $0.6146$ & $0.664$ \\ 
     \hline
     2 & -10.75 & $-9.25$  & $0.8$ & $0.58$ & $0.978$ \\ 
     \hline
     \hline
\end{tabular}
\caption{Amplification stages, demonstrated for TFIM with $J=1$, $g=1$, system sizes $L = 6$ and $L = 8$ (spectral gap in the cosine space: $\Delta_a \coloneqq a_0 - a_1 \approx 0.004$ and $\Delta_a \approx 0.003$). Between stages, Spectrum Profiling (Alg. \ref{alg:alg2}) was used in order to determine $(\lambda_{\text{LB}}, \lambda_{\text{UB}})$ of the next stage. The initial cut-off value $\mu^0$ was set to $0.95$. $p$ in the Table represents the success probability of resetting the ancilla qubit to $\ket{0}$. The initial state was randomized with an overlap of $\abs{\braket{\psi_0|\psi_{\text{init}}}} \approx 0.014$ and $0.002$ for $L=6$ and $L=8$ respectively. Compared to our method, repeating the first QETU layer ($i=0$) statically by the same number of times $M=3$ results in an overlap of $\approx 0.62$ and $\approx 0.35$, respectively.}
\label{tab:amplification}
\end{table}

\begin{table}[]
    \centering
    \text{Ground State Energy Estimation} \vspace{0.3cm}\\
    \begin{tabular}{||c c c c||} 
     \hline
     $p_{\text{Depolar}}$  & $\hspace{0.2cm} \text{DEM}$ \hspace{0.2cm}  & \hspace{0.2cm}  $\text{QCELS} \hspace{0.2cm} $ & \hspace{0.2cm}  $\text{RPE}$ \hspace{0.2cm} \\ [0.5ex] 
     \hline\hline
     $10^{-3}$ & $0.668$ & $0.25$ & $0.608$   \\ 
     $10^{-4}$ & $0.472$ & $0.0048$ & $9.2 \cdot 10^{-4}$   \\ 
     $10^{-5}$ & $0.579$ & $8.3 \cdot 10^{-4}$ & $9.7 \cdot 10^{-4}$   \\
     \hline
    \end{tabular}\\
    \caption{Absolute error in approximating the ground state energy of the TFIM Hamiltonian ($L=6$, $J=1$, $g=1$, $\lambda_0 \approx -7.727$) using Direct Expectation Value Measurement (DEM), QCELS \cite{qcels} and RPE \cite{rpe}. We present results from ground state preparation through Adaptive Finer Filtering (AFF) (parameters kept same as in Table \ref{tab:amplification}). $p_{\text{Depolar}}$ represents the local, two-qubit depolarizing probability, whereas single-qubit depolarizing probability was taken as $\frac{p_{\text{Depolar}}}{10}$. DEM was conducted with $10^4$ samplings on all qubits. RPE and QCELS were conducted with $10^4$ samplings on the ancilla qubit per each targeted phase. Simulations ran up to maximal simulation time $T_{\text{Max}} = 2^7$. }
    \label{tab:benchmark}
\end{table}

\begin{figure*}
\centering
\includegraphics[scale=0.5]{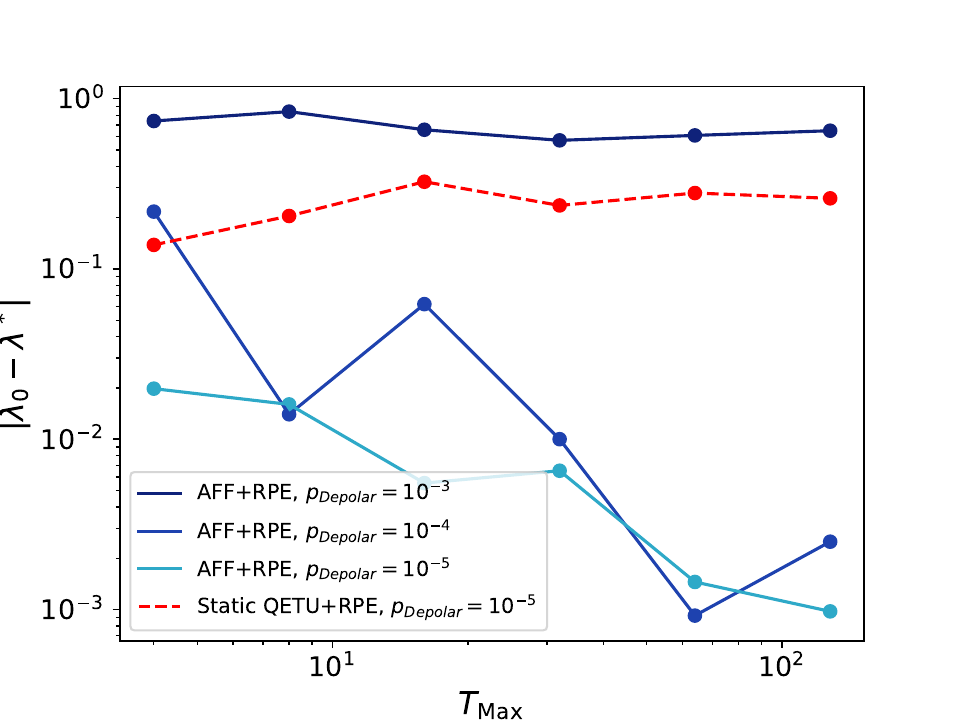}
\includegraphics[scale=0.5]{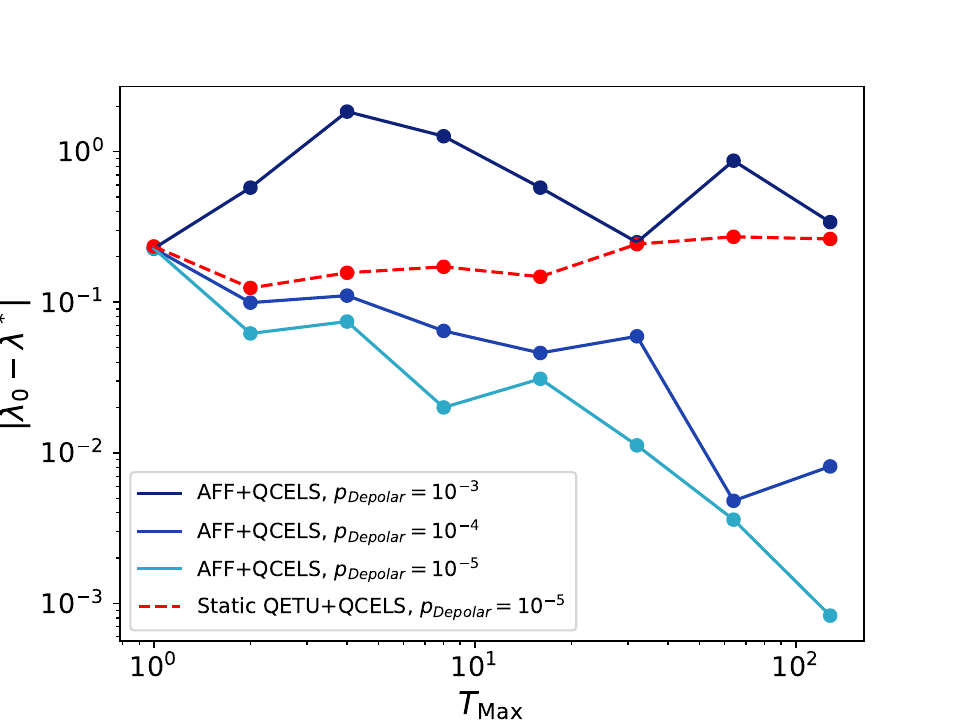}
\caption{Ground state energy estimation using Adaptive Finer Filtering (AFF), combined with Robust Phase Estimation \cite{rpe} (left) and QCELS \cite{qcels} (right). For comparison, the performance of statically applying QETU with the same number of repetitions $M=3$ and combining it with RPE / QCELS is plotted alongside in each graph. Simulations were conducted with $10^4$ (of which around $1\%$ corresponds to successful amplification) shots on the ancilla qubit per real and imaginary part of the targeted phase. The TFIM model with system size $L=6$ and system parameters $J=1, g=1$ (Spectral gap $\Delta \approx 0.263$) was used. Here, $p_{\text{Depolar}}$ represents the local depolarizing probability of two-qubit gates. The depolarizing probability of single-qubit gates is taken as $p_{\text{Depolar}}/10$.}
\label{fig:energy_plots}
\end{figure*}

To test the success of our algorithm in estimating the ground state energy, we combine Adaptive Finer Filtering with Robust Phase Estimation (RPE), developed by Ni et al.~\cite{rpe}, Direct Expectation Value Measurement as proposed by Dong et al. \cite{qetu} and Quantum Complex Exponential Least Squares (QCELS) algorithm as developed by Ding et al.\cite{qcels}. Results of the simulations with RPE and QCELS are demonstrated in Figure \ref{fig:energy_plots}. For further details on the implementation of these algorithms, you can refer to Appendix \ref{sec:B}, \ref{sec:C}, and \ref{sec:D}. The best results acquired through simulations are displayed in Table~\ref{tab:benchmark}.

\begin{figure}
\centering
\includegraphics[scale=0.45]{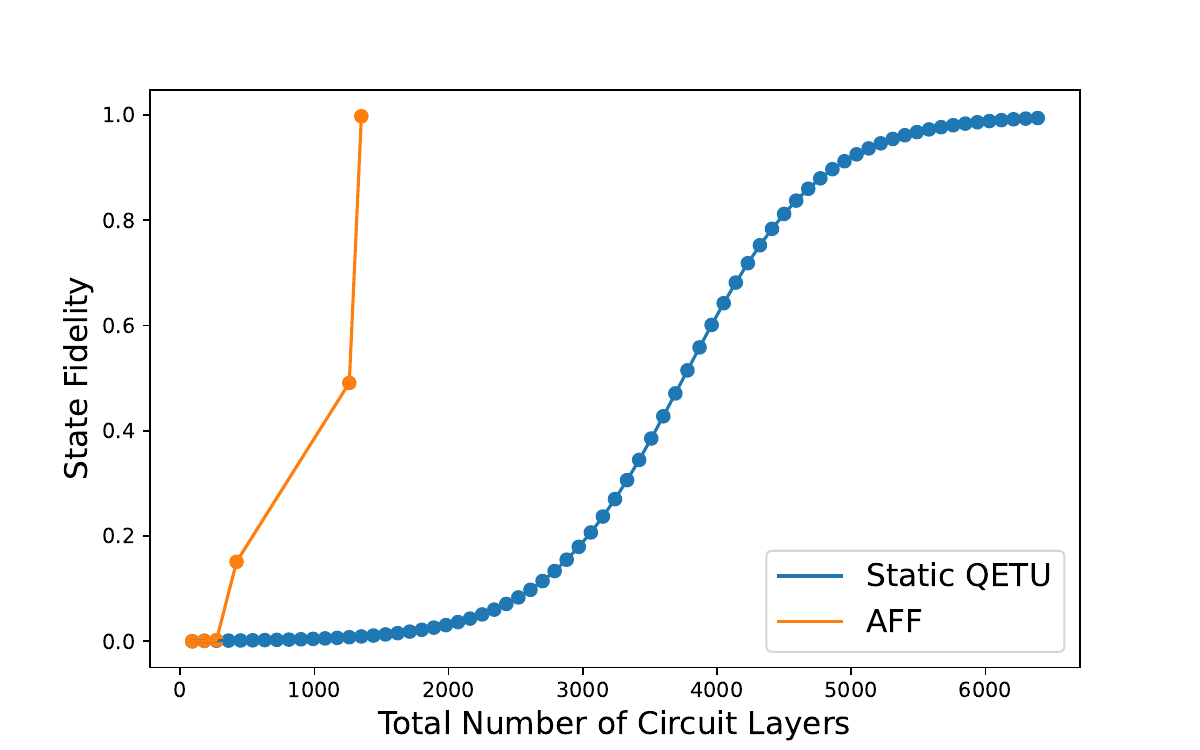}
\caption{Ground state preparation for TFIM with system size $L=10$, for small initial overlap $\abs{\braket{\psi|\psi_0}} \approx 10^{-7}$. For the static application of QETU, we assume we have access to a perfect cut-off value $\mu \in (a_1, a_0)$ (this strict assumption can be alleviated for AFF). ''Total Number of Circuit Layers'' represents the number of layers in the brick wall layout, used in RQC-Opt optimization to encode the time evolution operator, scaled by the polynomial degree $\eta = 30$. RQC-Opt needs only up to 7 layers per time evolution block if time coefficient $t<1$. For the case $t>1$, we split $dt=\frac{t}{n}$, such that $dt<1$ and repeat the circuit $n$ times. It is remarkable that in the fifth filtering stage of the AFF approach, we use the largest number of circuit layers, but to reach $\abs{\braket{\psi_0|\psi_f}} \approx 1$, we need to apply the first filtering (with lowest $t$ coefficient) again. This dampens the higher excited states once more and demonstrates the relative amplification between the ground and first excited states by the sixth stage of AFF.}
\label{fig:L10_plot1}
\end{figure}

For the system size $L=10$, we investigated a particularly challenging scenario, where $\sum_j \abs{\braket{\psi_j | \psi}} \approx 10^{-7}$ for $\ket{\psi_j}$ being the first $2^4$ low energy eigenstates. This increases the number of filtering stages $M$, which might cause unwanted amplification of higher energy states. This can be solved by applying the same QETU circuit from the first filtering stage once more at the end. An example demonstration of this phenomenon is given in Fig.~\ref{fig:L10_plot1}.

The need for a large enough initial overlap $\braket{\psi_0|\psi_{\text{init}}}$ still persists, so that the success probability of amplification is sensibly high. For example, the end result demonstrated in Table~\ref{tab:amplification} suggests a cumulative probability of around 0.75\% that the end state after the third amplification stage can be achieved. This low probability comes from the fact that the initial state was randomized and, hence, did not have a high overlap with low-energy states. Bearing this limitation in mind, we ran our simulations with a realistic, limited amount of shots ($10^4 \text{ to } 10^5$). We observe that most ($\approx 99\%$) of the total capacity to perform measurements on the ancilla qubit, results in the ancilla qubit being reset to $\ket{1}$ during one of the amplification stages. Despite this, we can still achieve the results demonstrated in Fig.~\ref{fig:energy_plots} by employing RPE or QCELS.

Another important point is the total simulation time of the ground state preparation process. Although our approach decreases the maximal simulation time for $\Tilde{\epsilon} \ll 1$, it causes an increase in the total simulation time as we introduce intermediate Hadamard sampling between amplification layers to perform the spectrum profiling. In practice, doing four samplings by setting the number of Fourier terms to $D=7$ delivers good results.

\section{Conclusion and Outlook}
As quantum computers are becoming able to accommodate more qubits and have the potential to outperform classical computers for certain tasks, quantum algorithms should optimize their scalability. With our proposed approach, we achieve a ground state preparation circuit, whose depth scales linearly with the inverse spectral gap. This scaling is similar to the state-of-the-art approaches such as the ones presented in \cite{Lin_2020, ge2018fastergroundstatepreparation}. What makes our approach particularly useful is to give a concrete recipe to tackle the practical challenges of using eigenspace filtering for ground state preparation, such as: the uncertainty of choosing an accurate cut-off value $\mu$ and decreasing spectral gap for large systems. This is advantageous in terms of reducing the deepest circuit that has to be accommodated by the quantum hardware in order to run this algorithm, however our approach increases the "total simulation time" as it increases the number of steps that have to be carried out. This is mainly due to the necessary, intermediate step of spectrum profiling. However, this does not increase the longest circuit depth of the algorithm.

The reliability and efficiency of our method can be improved by employing a quantum oracle to prepare an initial state for the amplification stages with an overlap with the ground state larger than a certain threshold. Methods such as single ancilla-Lindbladian evolution \cite{lindbladian} and Adiabatic Evolution \cite{adiabatic, keever2023adiabatic} can be tested as a pre-processing stage for the approach presented in this work. Another exciting direction could be employing classical algorithms such as DMRG \cite{DMRG1, DMRG2} or Coupled Cluster \cite{CC} to obtain an initial state. This state can then be prepared by the quantum hardware through known methods such as Ref.~\cite{mottonen2004transformation} proposed by Mottonen et al., by unitary dilation of the matrix product state tensors \cite{Smith2022, Malz2024}, or sparsification methods \cite{kiss2024early}.

Methods of efficiently implementing time evolution blocks with large time coefficients should be investigated \cite{classically_opt} because the RPE algorithm we apply at the end of the amplification layers promises a high potential of enhancing the noise tolerance if we can prevent the depth of time evolution circuit from scaling linearly with increasing time coefficient.

On top of depolarizing noise, the effects of other noise models, such as amplitude damping \cite{Nielsen_Chuang_2010, aziz2022thermal}, dephasing error \cite{Chuang_1995,Barenco_1996,Unruh_1995}, coherent errors in the form of over-rotation \cite{tan2023error_1, tan2023error_2} and measurement noise \cite{Wudarski_2023, Wudarski_2023_2, nachman2020unfolding} can be further investigated.

Possibly enhancing our method by combining it with other error correction methods, such as the noise estimation \cite{Urbanek_2021}, randomized compiling \cite{Random_Compiling} and using a coherent recovery sequence \cite{tan2023error_1, tan2023error_2} are interesting directions for future work.

\section*{Author Contributions}
The authors confirm their contribution to the paper as follows: 
Study conception and design: E.~Karacan, C.~B.~Mendl; Implementation and Simulation: E.~Karacan; Analysis and interpretation of results: E.~Karacan, C.~B.~Mendl; Draft Manuscript Preparation: E.~Karacan, Y.~Chen, C.~B.~Mendl. All authors reviewed the results and approved the final version of the manuscript.

\begin{acknowledgments}
We would like to thank L. Lin, Y. Dong and Z. Ding for insightful discussions during the conceptual development and implementation of this project.
Y.~Chen acknowledges funding by the Munich Quantum Valley section K7 (QACI), which is supported by the Bavarian state government with funds from the Hightech Agenda Bayern Plus.
\end{acknowledgments}
\vspace{1cm}
\onecolumngrid
\appendix
\section{Circuit Depth Scaling for Static Repetition of QETU}
\label{sec:A}
In this section, we derive the result presented in Proposition~\ref{prop:static_depth}. We will assume the non degenerate case for the ground and first excited states, which fits most of the use cases. For the degenerate case one can show the same end result. Firstly, we can write the relative amplification, defined in \ref{def:relative_ampf} as the following:
\begin{equation}
    A = \frac{F(a_0)}{F(a_0 - \Delta_a)}
\end{equation}
where $\Delta_a \coloneqq a_0 - a_1$ is the spectral gap in the cosine space. We then make the following approximation for the relative amplification (when static repetition approach is employed):
\begin{equation}
    \label{eq:static_ampf_approx}
    A_{\text{ST}} \approx \left(\frac{1-(a_0+1)^{-\eta}}{1-(a_0+1-\Delta_a)^{-\eta}}\right)^{\gamma}
\end{equation}
where $\gamma$ is the number of consecutive, static repetitions of the QETU circuit. Here, the polynomial $F(a)$ of degree $\eta$ is approximated through: 
\begin{equation}
    \label{eq:f_approx}
    F(a) \approx 1-(a+1)^{-\eta}.
\end{equation}
This approximation assumes the pessimistic scenario that $a_1 > \mu$, indicating that $F(a)$ is concave in the region around $[a_1, a_0]$. For the convex case, one can take $F(a) \approx \nu \left(\frac{a}{\mu}\right)^{\eta}$ where $\nu \in (0, 1)$. This does not change the end result, so we will continue with the concave assumption. To further simplify $A_{\text{ST}}$, we use Maclaurin series expansion:
\begin{equation}
    \label{eq:static_approx2}
    A_{\text{ST}} \approx \left(\frac{1-(1-\eta a_0 + \frac{\eta (\eta +1)}{2} a_0^2 + \dots)}{1-(1-\eta (a_0 - \Delta_a) + \frac{\eta (\eta+1)}{2}(a_0 - \Delta_a)^2 + \dots )}\right)^{\gamma}.
\end{equation}
Resulting from $a_0, a_0-\Delta_a < 1$, we further simplify the expression by ignoring the higher order terms as:
\begin{equation}
    \label{eq:static_approx3}
    A_{\text{ST}} \approx \left( \frac{a_0}{a_0 - \Delta_a} \right)^{\gamma}
\end{equation}
This approximation results in the following expression for $\gamma = \gamma_{\text{ST}}$ and $A_{\text{ST}} = \frac{a_0}{\Delta_a \Tilde{\epsilon}}$ (as given in \eqref{eq:relative_ampf_cond}):
\begin{equation}
    \gamma_{\text{ST}} \approx \frac{\log(1/A_{\text{ST}})}{\log(1-\frac{\Delta_a}{a_0})} =  \frac{\log(\frac{\Delta_a \Tilde{\epsilon}}{a_0})}{\log(1-\frac{\Delta_a}{a_0})}
\end{equation}
We use the Taylor expansion of $\log(1+x) = x - \frac{x^2}{2} + \dots$ and obtain the following result:
\begin{equation}
    \label{eq:unscaled_result}
    \gamma_{\text{ST}} \approx \frac{\log(\frac{\Delta_a \Tilde{\epsilon}}{a_0})}{-\frac{\Delta_a}{a_0} -0.5 \left(  \frac{\Delta_a}{a_0}\right)^2 + \dots} \approx  \frac{a_0}{\Delta_a} \cdot \log(\frac{a_0}{\Delta_a \Tilde{\epsilon}})
\end{equation}
We further aim to avoid the dependence of the above expression on $\Delta_a$ by replacing it with an expression dependent on spectral gap in the eigenvalue space $\Delta$. For this purpose, we derive a relation between $\Delta_a$ and $\Delta$:
\begin{align}
    \label{eq: delta_scaling}
    \Delta_a  &= \cos\left(\frac{\Tilde{\lambda}_0}{2}\right) - \cos\left(\frac{\Tilde{\lambda}_1}{2}\right)\\
    &= -2 \sin\left(\frac{\Tilde{\lambda}_0 + \Tilde{\lambda}_1}{2}\right) \sin\left(\frac{\Tilde{\lambda}_0 - \Tilde{\lambda}_1}{2}\right)\\
    &= 2 \sin\left(\frac{\Tilde{\lambda}_0 + \Tilde{\lambda}_1}{2}\right) \sin\left(\frac{\Delta}{2}\right)\\
    &\approx \frac{\Delta (\Tilde{\lambda}_1 + \Tilde{\lambda}_0)}{2} \hspace{0.2cm} = \hspace{0.2cm} \mathcal{O} (\Delta) 
\end{align}
where we use the special case for our (linear transformed) Hamiltonian  $\Tilde{H}$, that the spectrum is compressed into the $(0, \pi)$ range. To ensure this property, we apply the linear transformation given in \eqref{eq:linear_trafo} as $\Tilde{H} = \frac{\pi}{\Lambda} (H - \lambda_{\text{LB}}I)$, which (in practice) results in $|\Tilde{\lambda}_0|, |\Tilde{\lambda}_1| \ll 1$, justifying the approximation above.\\
By combining this relation and our result from \eqref{eq:unscaled_result}, we finally achieve the following scaling:
\begin{equation}
    \gamma_{\text{ST}} = \mathcal{O} (\Delta^{-1} \log(\Delta^{-1} \Tilde{\epsilon}^{-1})).
\end{equation}

\section{Circuit Depth Scaling for Adaptive Finer Filtering}
\label{sec:proof2}
In this section, we show that Algorithm \ref{alg:alg1} achieves circuit depth scaling $\gamma_{\text{AFF}}$ as given in Eq.~\eqref{eq:aff_depth}.
Unlike the repetition based, static implementation, whose scaling we have derived in Appendix \ref{sec:A}, Adaptive Finer Filtering relies on the stretching of the Hamiltonian. We will first show that the Objective \ref{def:gsa} can be achieved through the scaling given in Eq.~\eqref{eq:aff_depth}, then comment on whether fulfilling Objective \ref{def:gsa} still gives us a one-to-one correspondence with the standard ground state preparation task, that we defined in Obj. \ref{obj:gsp}.

Similar to the approximation given in Eq.~\eqref{eq:static_ampf_approx}, we approximate the amplification that can be achieved by Adaptive Finer Filtering as:
\begin{equation}
    \label{eq:aff_approx}
    A_{\text{AFF}} \approx \left(\frac{1-(a_0+1)^{-\eta}}{1-(a_0+1-\Tilde{\gamma} \Delta_a)^{-\eta}} \right)
\end{equation}
where the same approximation for $F(a)$ as in Eq.~\eqref{eq:f_approx} is used and $\Tilde{\gamma} < \frac{a_0}{\Delta_a}$ is the ''stretch parameter'' in the cosine space, that is related to the spectral gap as:
\begin{equation}
\begin{split}
    \label{eq:relation_gammas}
    \Tilde{\gamma} \Delta_a &= \cos\left(\frac{\gamma \Tilde{\lambda}_0}{2}\right) - \cos\left(\frac{\gamma \Tilde{\lambda}_1}{2}\right)\\
    &= -2 \sin \left( \frac{(\Tilde{\lambda}_0 + \Tilde{\lambda}_1)\gamma}{2} \right) \sin \left(\frac{(\Tilde{\lambda}_0 - \Tilde{\lambda}_1)\gamma}{2} \right).
\end{split}
\end{equation}
Here $\gamma$ is the stretch parameter in the eigenvalue space, hence $\gamma$ is also what we take as the circuit depth (in accordance to Def.~\ref{def:circuit_depth}). 

We first consider the range of $\Tilde{\gamma}$, in order to achieve the targeted relative amplification.
By Eq.~\eqref{eq:aff_approx} and using the same steps as in \eqref{eq:static_approx2} and \eqref{eq:static_approx3}, we obtain:
\begin{equation}
        A_{\text{AFF}}  \approx  \left(\frac{1-(1-\eta a_0 + \frac{\eta (\eta +1)}{2} a_0^2 + \dots)}{1-(1-\eta (a_0 - \Tilde{\gamma} \Delta_a ) + \frac{\eta (\eta+1)}{2}(a_0 - \Tilde{\gamma} \Delta_a )^2 + \dots )}\right) \hspace{0.3cm}\approx \hspace{0.3cm} \frac{a_0}{a_0 - \Tilde{\gamma} \Delta_a}
\end{equation}
This approximation results in the following expression for $\Tilde{\gamma} = \Tilde{\gamma}_{\text{AFF}}$ and $A_{\text{AFF}} = \frac{a_0}{\Delta_a \Tilde{\epsilon}}$ (as given in \eqref{eq:relative_ampf_cond}):
\begin{equation}
    1-\frac{\Delta_a \Tilde{\gamma}_{\text{AFF}}}{a_0} \hspace{0.1cm} = \frac{\Delta_a \Tilde{\epsilon}}{a_0}
\end{equation}
\begin{equation}
    \label{eq:gamma_tilde_result}
    \Tilde{\gamma}_{\text{AFF}} \hspace{0.1cm} = \frac{a_0}{\Delta_a} - \Tilde{\epsilon}
\end{equation}

\begin{figure}
\centering
\includegraphics[scale=0.5]{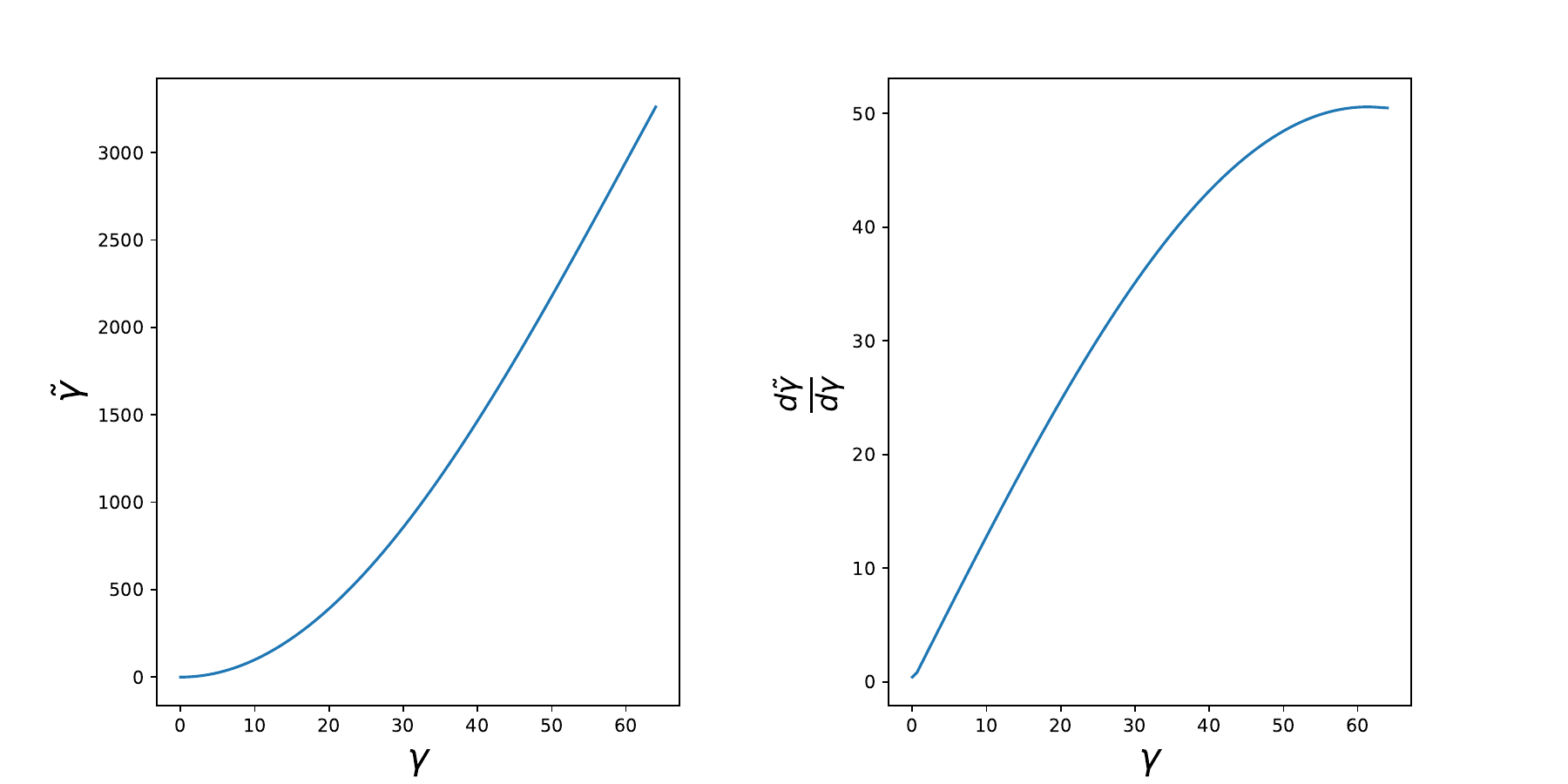}
\caption{Numerical simulation plots demonstrating the relation between $\Tilde{\gamma}$ and $\gamma$ according to \eqref{eq:relation_gammas}. Eigenvalues ($\Tilde{\lambda}_0, \Tilde{\lambda}_1$) are set to (0.01, 0.05) to match the spectrum of the linearly scaled Hamiltonian (as given in Eq.~\eqref{eq:linear_trafo}). We observe a quadratic scaling for small $\Tilde{\gamma}$ values, however scaling becomes linear in the range we operate ($\Tilde{A} = \frac{a_0}{\Delta_a \Tilde{\epsilon}}$ indicating $\Tilde{\gamma} \approx \frac{a_0}{\Delta_a} > 3000$).}
\label{fig:gamma_scaling}
\end{figure}

In order to derive the scaling for the circuit depth $\gamma$, we numerically simulate the relation between $\Tilde{\gamma}$ and $\gamma$ for $\Tilde{\gamma} \approx \frac{a_0}{\Delta_a}$. This is necessary because, unlike the assumptions made in Eq.~\eqref{eq: delta_scaling}, we cannot take the arguments of sine terms in Eq.~\eqref{eq:relation_gammas} to be small enough to justify a linear approximation of the sine function. As the result of the numerical simulation, we observe a linear scaling as:
\begin{equation}
    \label{eq:gamma_tilde_scaling}
    \Tilde{\gamma} \Delta_a = \mathcal{O} (\gamma \Delta)
\end{equation}
The example plots for the numerical simulation are presented in Fig. \ref{fig:gamma_scaling}. By using the result from \eqref{eq:gamma_tilde_result}, \eqref{eq:gamma_tilde_scaling} and $\Delta_a^{-1} \gg \Tilde{\epsilon}$, we finally obtain the scaling as:
\begin{equation}
    \gamma_{\text{AFF}} = \mathcal{O}(\Delta^{-1}).
\end{equation}

Hence, one single filtering stage of AFF requires a circuit depth that scales as $\mathcal{O}(\Delta^{-1})$, in order to achieve relative amplification $\Tilde{A}$ of the ground state (with respect to the first excitation), that satisfies Obj. \ref{def:gsa}. In practice, this does not indicate that with this one single filtering stage we can also fulfill Obj. \ref{obj:gsp} and obtain a good approximation of the ground state. The reason for this, is that Eq.~\eqref{eq:A_to_inf} does not directly hold anymore for a stretched Hamiltonian, as the spectrum cannot be mapped to the cosine space bijectively. 

That is why we require a set of ''initial filtering'' stages (the number of which is given as $M-1$ in Alg.~\ref{alg:alg1}) to eliminate excitations above the low energy sub-space. Each of these initial filtering stages stretch the Hamiltonian by a factor $\gamma_j = \mathcal{O}(\Delta_j^{-1})$, where  $\Delta_j \coloneqq a_0 - a_j, \hspace{0.1cm} (j > 1)$ is the $\text{j}^{\text{th}}$-higher spectral gap. Consequently, initial filtering stages bring additive factors to the circuit depth as: \begin{equation}
    \gamma_{\text{AFF}} = \mathcal{O}\left(\Delta^{-1} + \sum_{\Delta_j \in B} \Delta_j^{-1} \right)
\end{equation}
where $B$ is the subset of higher spectral gaps with number of elements $M-1$, each element of which corresponds to an initial filtering stage. 

If we assume that the spectrum is evenly spread out in the cosine space, and the $\eta, \mu$ parameters of each initial filtering stage are chosen such that we acquire $1/\ell$ of the spectrum after each filtering, we would need
\begin{equation}
    M = \ceil*{\log_{\ell}N}
\end{equation}
initial filtering steps in the worst case (scenario where the initial state has non-negligible overlaps with all excited states), where $N$ is the number of dimensions of the Hamiltonian. With the same assumptions as above, we can define the higher spectral gaps that make up the subset as:
\begin{equation}
    B = \{ \Delta_{\ceil*{N/\ell^j}} \}_{j=0}^{M-2}.
\end{equation}
For our circuit depth (as given in  Eq.~\eqref{eq:aff_depth}), we ignore these additive factors with the assumption of $\Delta \ll \Delta_j \hspace{0.2cm} \forall \Delta_j \in B$, which is the case for most of the Hamiltonians of interest.

\section{Direct Expectation Value Measurement}
\label{sec:B}
The most straightforward idea of extracting the ground state energy information from the prepared ground state is to directly conduct measurements on the ground state to approximate the probability distribution of the state vector. This works in cases where the expectation value of the energy can be decomposed into a combination of Pauli operations, which corresponds to calculating the expectation value of the ground state w.r.t.\ a different measurement basis. This approach inevitably faces a stochastic limit as approximating the probability distribution with high precision for large systems is impossible with a sensible amount of experiments. Hence, the lower bound of approximation error significantly increases with growing system size, even in the ideal, noiseless case.

For the TFIM Hamiltonian, this corresponds to:
\begin{equation}
\begin{split}
E_0 &= \bra{\psi_0} H_{\text{TFIM}} \ket{\psi_0} = -J\sum_{j=1}^{L-1} \bra{\psi_0} Z_j Z_{j+1} \ket{\psi_0} - g \sum_{j=1}^{L} \bra{\psi_0} X_j \ket{\psi_0}\\
&= -J\sum_{j=1}^{L-1} \sum_{\sigma_j=0}^1 \sum_{\sigma_{j+1}=0}^1 (-1)^{\sigma_j}  (-1)^{\sigma_{j+1}} \mathds{P}_Z (\sigma_j, \sigma_{j+1}| \psi_0) - g \sum_{j=1}^{L}  \sum_{\sigma_j=0}^1(-1)^{\sigma_j} \mathds{P}_X (\sigma_j| \psi_0)
\end{split}
\end{equation}
 where $\mathds{P}_X (\sigma_j| \psi_0)$ is the probability of measuring $j^{th}$ qubit as $\ket{\sigma_j}$ on the $X$-basis and $ \mathds{P}_Z (\sigma_j, \sigma_{j+1}| \psi_0)$ is the probability of measuring  $j^{th}$ and $(j+1)^{th}$ qubits as  $\ket{\sigma_j \sigma_{j+1}}$ on the computational basis ($Z$-basis).
 
The simulation of conducting measurements w.r.t.\ a different basis than the computational basis is performed by applying a Hadamard gate to all the qubits for the $X$-measurement, as a consequence of $X = H Z H$.

\section{Robust Phase Estimation (RPE)}
\label{sec:C}
The Robust Phase Estimation algorithm, as proposed by Ni et al.~\cite{rpe}, approximates the ground state energy from an end state whose overlap with the ground state is large enough ($>0.53$) digit by digit on the binary basis, by employing the same Hadamard circuit, as the one given in Fig. \ref{fig:hadamard_circuit}, whose time coefficient is set to $t=2^j$, where $j$ is the target digit the search aims to identify. By successively increasing $j$ and using the result from the predecessing search stage, RPE enables highly noise-resilient and robust approximation of the ground state energy, especially if the initial overlap is high $\approx 1$.\\
We start with a rough estimate $\theta_{-1}$, that can be acquired through methods such as DEM (Appendix \ref{sec:B}) and follow the given Algorithm \ref{alg:alg3}:
\begin{algorithm}
\caption{Robust Phase Estimation (RPE) \\(proposed by Ni et al.~\cite{rpe})}\label{alg:alg3}
\KwData{$\ket{\psi}, \hspace{0.1cm}  \theta_{-1},\hspace{0.1cm} J, \hspace{0.1cm} N_S$}

\For {$j=0, \dots , J-1$} {
    $Z_j \approx \braket{\psi|\e^{i2^jH}|\psi} $ \tcp{Acquired through the Hadamard test circuit in Fig.~\ref{fig:hadamard_circuit} with $N_S$ samples}
    $S_j \coloneqq \{\frac{2k\pi + argZ_j}{2^j} \}_{k=0, \dots, 2^j-1}$\\
    $\theta_j = \argmin_{\theta \in S_j} [\pi - |(\theta - \theta_{j-1}, \text{ mod }2\pi ) - \pi|] $

}
\Return $\theta_{J-1}$
\end{algorithm}

This method is remarkable due to its short circuit depth, overcoming the stochastic sampling limitations (Monte-Carlo noise), and high performance despite small sampling size $N_S \in [10^2, 10^3]$. This method also tolerates local depolarizing noise well. For further details, please refer to \cite{rpe}. 

\section{Quantum Complex Exponential Least Squares (QCELS) Algorithm }
\label{sec:D}
Another ground state energy estimation algorithm we use in our benchmarks is the Quantum Complex Exponential Least Squares Algorithm (QCELS), as proposed by Ding et al. \cite{qcels}.

In this Algorithm, the same Hadamard test circuit (Fig. \ref{fig:hadamard_circuit}), with the number of samples $N_S$, is employed to estimate a set of phases $\{Z_n\}_{n=0}^{N-1}$:
\begin{equation}
     Z_n  \approx e^{-i\lambda_0t_n}
\end{equation}
for different $t_n$ values: $t_n = n\tau_j$, where $\tau_j$ gets successively higher as $\tau_j = \tau \cdot 2^j$, for $j=0, \dots, J-1$. Here $J, N_S, \tau$ and $N$ are constants to be set depending on system parameters.

\begin{algorithm}
\caption{ Quantum Complex Exponential Least Squares (QCELS) Algorithm \\(proposed by Ding et al. \cite{qcels})}\label{alg:alg4}
\KwData{$\ket{\psi}, \hspace{0.1cm}  \lambda_{LB},\hspace{0.1cm}, \lambda_{UB}, \hspace{0.1cm} J, \hspace{0.1cm} N_S, \hspace{0.1cm} N, \hspace{0.1cm} \tau$}
\For {$j=0, \dots , J-1$} {
    $\tau_j \gets \tau \cdot 2^j$ \vspace{0.1cm} \\
    $\{Z_{j, n}\}_{n=0}^{N-1} \approx \{\braket{\psi|\e^{i n \tau_j H}|\psi}\}_{n=0}^{N-1} $ \tcp{Acquired through the circuit in Fig.~\ref{fig:hadamard_circuit} with $N_S$ samples} \vspace{0.1cm}
    $\theta_j^* \gets \argmin_{r \in \mathbf{C}, \theta \in [\lambda_{LB}, \lambda_{UB}]} \frac{1}{N} \sum_{n=0}^{N-1} |Z_{j, n} - r\e^{-i\theta n \tau_j}|^2 $\\ \vspace{0.25cm}
    $(\lambda_{LB}, \lambda_{UB}) \gets (\theta^*_j - \frac{\pi}{2\tau_j}, \theta^*_j + \frac{\pi}{2\tau_j})$
}
\Return $\theta_{J-1}$
\end{algorithm}

On this data set $\{Z_n\}_{n=0}^{N-1}$, we perform an exponential fit ($r \e^{-i\theta t_n}$) at each stage $j$ and choose the optimal angle parameter $\theta^*_j$ as the result of the current stage. Next stage $j+1$ continues the optimization within the bounds limited by $(\theta^*_j - \frac{\pi}{2\tau_j}, \theta^*_j + \frac{\pi}{2\tau_j})$ with an increased time step $\tau_{j+1} = 2\tau_j$. For the results displayed in Table~\ref{tab:benchmark}, we used $N=5$, $\tau=0.2$, $J=9$, $N_S=10^4$, so that the maximal simulation time $T_{\max}$ matches the $T_{\max}$ of RPE (Appendix \ref{sec:C}). This algorithm is summarized in Alg.~\ref{alg:alg4}.

\vspace{1cm}
\twocolumngrid
\bibliography{references.bib}

\end{document}